# Dispersions of Many-Body Bethe strings

Anup Kumar Bera[1,2,*], Jianda Wu[3,*], Wang Yang[4,*], Zhe Wang[5], Robert Bewley,[6] Martin Boehm,[7] Maciej Bartkowiak,[1] Oleksandr Prokhnenko,[1] Bastian Klemke,[1] A. T. M. Nazmul Islam,[1] Joseph Mathew Law,[8] Bella Lake[1,9,*]

[1] Helmholtz-Zentrum Berlin für Materialien und Energie, 14109 Berlin, Germany
[2] Solid State Physics Division, Bhabha Atomic Research Centre, Mumbai 400085, India
[3] Tsung-Dao Lee Institute&School of Physic and Astronomy, Shanghai Jiao Tong University, 800 Dongchuan Rd., Shanghai, 200240, China
[4] Stewart Blusson Quantum Matter Institute, 2355 East Mall Vancouver, BC, V6T 1Z4, Canada
[5] Institute of Physics II, University of Cologne, 50937 Cologne, Germany
[6] ISIS Facility, STFC Rutherford Appleton Laboratory, Harwell Oxford, Didcot OX11 0QX, United Kingdom
[7] Institut Laue-Langevin, 71 Avenue des Martyrs, 38042 Grenoble, France
[8] Hochfeld-Magnetlabor Dresden (HLD-EMFL), Helmholtz-Zentrum Dresden-Rossendorf, D-01314 Dresden, Germany
[9] Institut für Festkörperphysik, Technische Universität Berlin, 10623 Berlin, Germany
*Corresponding author. Email: akbera@barc.gov.in (A.K.B.), wujd@sjtu.edu.cn (J.W.), wang.yang@ubc.ca (W. Y.), bella.lake@helmholtz-berlin.de (B.L.)

**Complex bound states of magnetic excitations, known as Bethe string, were predicted almost a century ago to exist in one-dimensional quantum magnets[1]. The dispersions of the string states have so far remained the subject of intensive theoretical studies[2-7]. By performing neutron scattering experiments on the one-dimensional Heisenberg–Ising antiferromagnet $SrCo_2V_2O_8$ in high longitudinal magnetic fields, we reveal in detail the dispersion relations of the string states over the full Brillouin zone, as well as their magnetic field dependences. Furthermore the characteristic energy, the scattering intensity and linewidth of the observed string states exhibit excellent agreement with our precise Bethe Ansatz calculations. Our results establish the important role of string states in the quantum spin dynamics of one-dimensional systems, and will invoke studies of their dynamical properties in more general many-body systems.**

In a solid, states of matter emerge from many-body interactions of the electrons and nuclei that constitute the lattice of the solid characterized by specific dimensionalities and symmetries. To identify a new state of matter, it is insufficient to characterize only its ground-state orders, but necessary also to reveal the unique energy-momentum relations of featured excitations resulting from the many-body interactions. This is not only because different interactions can lead to the same ground-state order, but also due to the fact that an ordered ground state can be absent, as is the case for one-dimension (1D) systems, even at zero temperature. Further challenges arise in the unambiguous identification of a new state of matter due to the difficulties of precisely solving a many-body problem and/or of properly realizing a solvable model in a real material. We overcome these difficulties in a realization of the exactly solvable model of the 1D spin-chain in $SrCo_2V_2O_8$. Our study reveals the characteristic dispersion relations and thereby identifies unambiguously the excited states of many-body Bethe strings.

In 1931 Hans Bethe attempted to solve the many-body interaction model of the 1D isotropic Heisenberg spin-1/2 chain, and predicted the existence of two-magnon bound states[1]. The systematic ansatz he introduced has been extended to predict many-body string states (containing complex multiple magnon bound states) in the generalized 1D anisotropic Heisenberg-Ising (or XXZ) model[2-8]. However, compared with the excitations of spinons that were found to govern the spin dynamics in zero magnetic field[9,10], the string states were deemed undetectable to conventional experimental techniques due to their negligible spectral weight[11,12]. Nonetheless, a recent terahertz spectroscopic study in high magnetic fields, focusing on the zero-momentum-transfer point (the Γ point in reciprocal space), provided evidence for the string states in a Heisenberg-Ising chain[13]. On the other hand, so far the dispersion relations of the string states remain experimentally unexplored leaving the identification incomplete. The recent development of facilities for high-field inelastic neutron scattering and the availability of high quality, large single crystals[14] now make such an experimental attempt possible. Here we report inelastic neutron scattering on the 1D Heisenberg-Ising spin-chain compound $SrCo_2V_2O_8$ in its field-induced quantum critical phase at unprecedentedly high longitudinal magnetic fields up to 25.9 T. Our experimentally obtained dispersion relations, scattering intensity, as well as lineshapes as a function of magnetic field are compared in detail to precise Bethe-Ansatz calculations. The excellent agreement between experiment and theory allows the unambiguous identification and full characterisation of the complex two- and three-string states, completing the verification of the many-body Bethe strings.

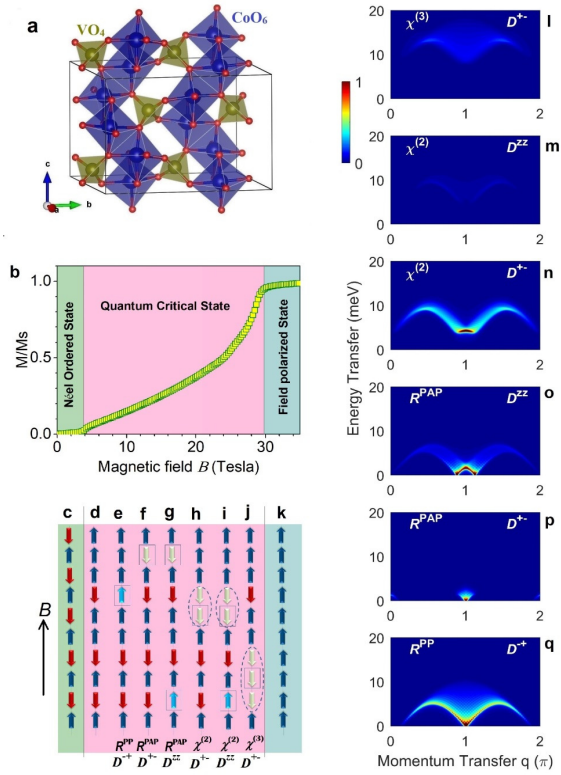

**Figure 1| Crystal structure and magnetic phase diagram of $SrCo_2V_2O_8$ along with the theoretical spectra in the critical regime. a,** The tetragonal crystal structure of $SrCo_2V_2O_8$ contains screw chains of edge-sharing $CoO_6$ octahedra running parallel to the $c$ axis separated by nonmagnetic $VO_4$ tetrahedra. Only half of the unit cell (2 chains) is presented for clarity, and the nonmagnetic $Sr^{2+}$ atoms are also omitted. The unit cell dimensions are shown by the gray lines. For more crystal structure details, see the supplementary information (SI)[14]. **b,** The magnetization curve of $SrCo_2V_2O_8$ showing the relative magnetization $2m=M/M_S$ as a function of applied magnetic field $B$ along the $c$ axis at temperature $T$= 1.5 K. Three magnetic phases (Néel-ordered state for $B < B_C = 3.8$ T, quantum critical state for $B_C < B < B_S$ and field-polarized states for $B > B_S = 28.7$ T) are indicated by the green, pink and blue shading, respectively. **c,** The schematic representations of magnetic Néel-ordered state (NAF) (total spin-$z$ quantum number $S_Z^T=0$), and **k,** field-polarized (FP) state ($S_Z^T=N/2$), where, $N$ is the total numbers of spins. The direction of the applied longitudinal field is shown by the thin black arrow on the left. **d,** The ground state at an arbitrary field value $B'$ in the intermediate quantum critical regime $B_C < B < B_S$. The ground state with $S_Z^T=N/2-n$ has $n$ number of flipped spins (red arrows) with respect to the FP state. **e-j,** Schematic representations of excitation modes in the critical regime for the field $B'$; psinon–psinon ($R^{PP}$), psinon–antipsinon ($R^{PAP}$), two-string ($\chi^{(2)}$) and three-string ($\chi^{(3)}$) where the excited states are created by flipping spins with respect to (**d**). The flipped spins are indicated by rectangular boxes. The 2-string and 3-string states which contain bound states formed by two and three magnons, respectively, (shown by ovals) move as a single entities in the chain; in contrast to the psinon–(anti)psinon pairs that can propagate throughout the chain without forming bound states. The dynamical structure factors (DSFs) channels [with transverse ($D^{+-}$ and $D^{-+}$) and longitudinal ($D^{ZZ}$) polarizations] in which the modes appear are also stated. **l-q,** The dispersion relations and intensities for the $\chi^{(3)}$, $\chi^{(2)}$, $R^{PAP}$ and $R^{PP}$ excited states in the $D^{+-}$, $D^{-+}$ and $D^{ZZ}$ channels, calculated by the Bethe Ansatz method for the Hamiltonian of $SrCo_2V_2O_8$ ($J = 3.55$ meV and $\Delta = 2$) with magnetization $2m$ =0.12 corresponding to an applied longitudinal magnetic field of $B = 9$ T.

$SrCo_2V_2O_8$ is an excellent realization of the paradigmatic spin-interaction model known as the 1D spin-1/2 *XXZ* antiferromagnet. The magnetic $Co^{2+}$ ions have quantum spin-½ (S=1/2) angular momentum and are arranged in chains of edge-sharing $CoO_6$-octahedra providing a close approximation the Hamiltonian

$$H = J \sum_i (S_i^x S_{i+1}^x + S_i^y S_{i+1}^y + \Delta S_i^z S_{i+1}^z) - g\mu_B B \sum_i S_i^z \quad (1)$$

which considers the components $S_i^\alpha$ ($\alpha = x, y, z$) of a spin angular momentum operator, $S_i$ ($S = 1/2$) at site $i$ on a 1D chain. The positive $J$ denotes the dominant nearest-neighbor antiferromagnetic exchange coupling along the crystallographic $c$ direction (Fig. 1a) as realised by the crystal structure of $SrCo_2V_2O_8$[14]. $\Delta > 1$ is an Ising anisotropy parameter where the easy-axis (z-axis) is also along the $c$ direction, as confirmed by magnetization and neutron diffraction experiments. $B$ is an applied longitudinal magnetic field along the easy axis, while $g$ and $\mu_B$ are the Landé $g$-factor and the Bohr magneton, respectively. In zero field, a Néel-type collinear antiferromagnetic order (NAF) is stabilized below $T_N = 5$ K[15] due to weak interchain couplings [not included in Eq. (1)]. In the applied field, the NAF state is suppressed above $B_c = 3.8$ T and the system enters a gapless quantum critical state before reaching a field polarized ferromagnetic (FP) state at $B_S = 28.7$ T where all spins point along the field direction (Fig. 1b). The NAF and FP states can be illustrated by the antiparallel (Fig. 1c) and parallel (Fig. 1k) alignments of neighboring spins corresponding to a total spin-$z$ quantum number of $S_Z^T = 0$ and $N/2$, respectively, where $N$ is the total number of spins in a chain. In zero field the fundamental excitations in $SrCo_2V_2O_8$ are spinons[16], particles with fractional quantum spin $S=1/2$, which are fundamentally different from the $S = 1$ magnons or spin-wave excitations of conventional 3D magnets[17-19]. In the NAF state, the coupling between the spin chains leads to spinon confinement[16,20-22], with analogies to quark confinement in quantum chromodynamics[23-25]. In the critical regime ($Bc < B < Bs$), a representative ground state is illustrated for an arbitrary value of $S_Z^T$ (Fig. 1d). The dynamical spin structure factor (DSF) in this regime consists of the pair excitations psinon–psinon ($R^{PP}$) and psinon–antipsinon ($R^{PAP}$), as well as the two-string ($\chi^{(2)}$) and three-string ($\chi^{(3)}$) states which contain many-body bound states of two- and three-magnons, respectively (Fig. 1e–j)[26].

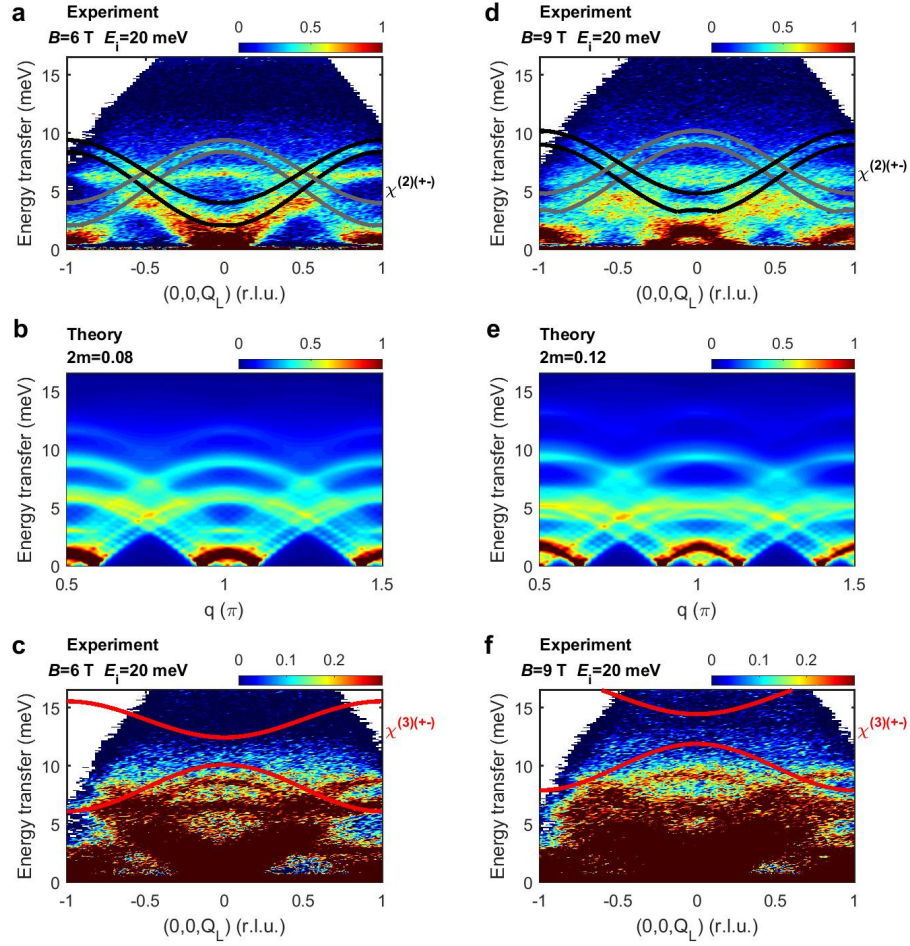

**Figure 2| Dispersion relations of Bethe strings in the energy and momentum space. a, c, d** and **f**, False color maps of the INS intensity measured on LET ISIS, UK at $T$=0.08 K under longitudinal magnetic fields of $B$= 6 (a and c) and 9 T (d and f) using a fixed incident energy $E_i$ = 20 meV, projected on the spin-chain direction (c axis). To gain intensity, the data were integrated over $Q_K$= -3.5–0 r.l.u. for **a** and **d;** and over $Q_H$ = -4–0, $Q_K$ = -7.0– -2.5 r.l.u. for **c** and **f**, respectively. The intensities are denoted by the colors, as indicated by the scales at the top. The black and gray lines in (a) and (d) show the dispersion boundaries of the 2-string excitations which are shifted by $Q_{L=\pm1}$ r.l.u, respectively, due to zone folding (see SI for details[14]). For (c) and (f), only one set of boundaries for the 3-string continuum is shown for clarity. The $R^{PAP}$ and $R^{PP}$ modes appear below ~ 10 meV. **b** and **e,** The excitation spectra for the one-dimensional spin-1/2 antiferromagnetic Heisenberg–Ising model which are calculated by Bethe ansatz with an exchange interaction $J$ = 3.55 meV, g-factors $g_\parallel$ = 5.5 and $g_\perp$ = 2.79, and Ising anisotropy $\Delta$ = 2.0 for $2m$ = 0.08 and 0.12 corresponding to applied fields of 6 and 9 T, respectively. The screw chain structure factor of $SrCo_2V_2O_8$, the $Co^{2+}$ form factor, the anisotropic Lande g-factor, and the polarization factors have been considered for the comparison with the experimental data.

The theoretical dispersion relations and the intensities of the $\chi^{(3)}$, $\chi^{(2)}$, $R^{PAP}$ and $R^{PP}$ excitations calculated by the Bethe-ansatz[26-28] for $J$ = 3.55 meV, $\Delta$ = 2 [Ref [16]] and $2m$ =0.12 which corresponds to $SrCo_2V_2O_8$ in a longitudinal magnetic field of $B$=9 T (where $2m=M/M_S$ is the relative magnetization and is a function of an applied magnetic field) are illustrated in Fig. 1l-q**.** All the modes form continua in the energy-momentum space. The string states ($\chi^{(3)}$ and $\chi^{(2)}$) are gapped throughout reciprocal space, whereas, $R^{PAP}$ and $R^{PP}$ are gapless at specific wavevectors. The string states $\chi^{(3)}$ and $\chi^{(2)}$ appear at high ($E = \hbar\omega$ > ~ 8 meV) and intermediate ($E$ > ~ 4 meV) energy ranges, respectively. The $R^{PAP}$ and $R^{PP}$ modes are found below 7 meV. All these modes involve fluctuations perpendicular to the easy axis and thus appear in the transverse dynamic structure factor (DSF) channels ($D^{+-}$ and $D^{-+}$), while the longitudinal channel ($D^{zz}$) which consists of fluctuations parallel to the easy axis is dominated by the $\chi^{(2)}$ and $R^{PAP}$ modes (Figs. 1m and 1o). For $\chi^{(2)}$, most of the intensity appears in the transverse $D^{+-}(q, E)$ channel with an M-shaped continuum (Fig. 1n). Around the zone center ($q= \pi$), the lower boundary of the continuum forms a W-shaped curve with field-dependent minima at $q = \pi \pm 2m\pi$. A similar W-shaped lower boundary is also evident for $R^{PAP}$ in the $D^{zz}$ channel (Fig. 1o). The spectral weight of all the other modes is heavily concentrated around $q=\pi$ except $\chi^{(3)}$ whose intensity is very weak and is distributed over all wavevectors (Fig. 1l).

We performed inelastic neutron scattering experiments (INS) to measure the dynamical structure factors of $SrCo_2V_2O_8$. Figure 2 depicts the results measured at $B = 6$ and 9 T which determine the experimental dispersion relations of the string states as well as of the psinon–psinon and psinon–antipsinon modes. Figures 2a and 2d show the full spectrum as a function of energy and wavevector. Obviously, the experimental data are more complex than the simple sum of the Bethe Ansatz results as depicted in Fig 1l-q. This is because the four-fold screw chain structure of $SrCo_2V_2O_8$ (Fig. 1a) leads to Brillouin-zone folding where four copies of the spectrum of a straight chain, each shifted successively by $q=\pi/2$, are observed simultaneously (see SI for more details[14]). This effect allows us to probe the dispersion relations of all the modes over the full Brillouin zone by only measuring the experimental neutron scattering pattern over $-1 \leq Q_L$ (r.l.u.) $\leq +1$ where $Q_L$ is the reciprocal $c$ lattice vector of $SrCo_2V_2O_8$. For a direct comparison between the experiment and theory, the zone-folding effects are applied to the Bethe Ansatz DSFs, and presented in Figs. 2b and 2e, which correspond to 6 and 9 T, respectively.

The most salient features are the dispersion relations of the 2- and 3-string states which appear in the intermediate and high energy regions (viz., 4–10.5 meV and 7.5–18 meV under $B=9$ T), respectively. The boundaries of the 2-string (Figs. 2a and 2d) and 3-string (Figs. 2c and 2f) continua are plotted over the data using black and red lines, respectively. Two sets of boundaries are shown for 2-string because of the zone folding. The dispersions of the 2-string continua are clearly visible in Fig 2a and 2d, while the intensity of the weaker 3-string states becomes visible on altering the intensity scale (Figs. 2c and 2f), and accounts for all the signal observed above 10 meV where the other modes are absent. The $R^{PAP}$ and $R^{PP}$ modes are also clearly visible. An energy cut through the data measured at $B=9$ T at the zone centre ($Q_L=0$) (Fig. 3) reveals all the transverse modes (fluctuating perpendicular to the easy $c$ axis) $R_{\pi/2}^{PP}$ ($q = \pi/2$), $\chi_\pi^{(2)}$ ($q = \pi$) and $\chi_{\pi/2}^{(3)}$ ($q = \pi/2$) as reported by THz spectroscopy[13]. In addition, the INS data also yields the first observations of the $\chi_{\pi/2}^{(2)}$ ($q = \pi/2$), $R_\pi^{PAP}$ ($q = \pi$), and $R_{\pi/2}^{PAP}$ ($q = \pi/2$) modes. The mode $R_\pi^{PAP}$ and $R_{\pi/2}^{PAP}$ originate from the longitudinal spin fluctuations (parallel to the easy $c$ axis) that are invisible to THz spectroscopy[13]. Cuts at other reciprocal lattice wavevector $Q_L=0.25$, 0.5 and 1.0, reveal a similarly rich spectrum (Fig. 3). The comparison to the theoretical results are excellent overall, while the slight difference at some transfer energies could be ascribed to effects of the interchain couplings which are small and have not been considered in the Bethe Ansatz calculations.

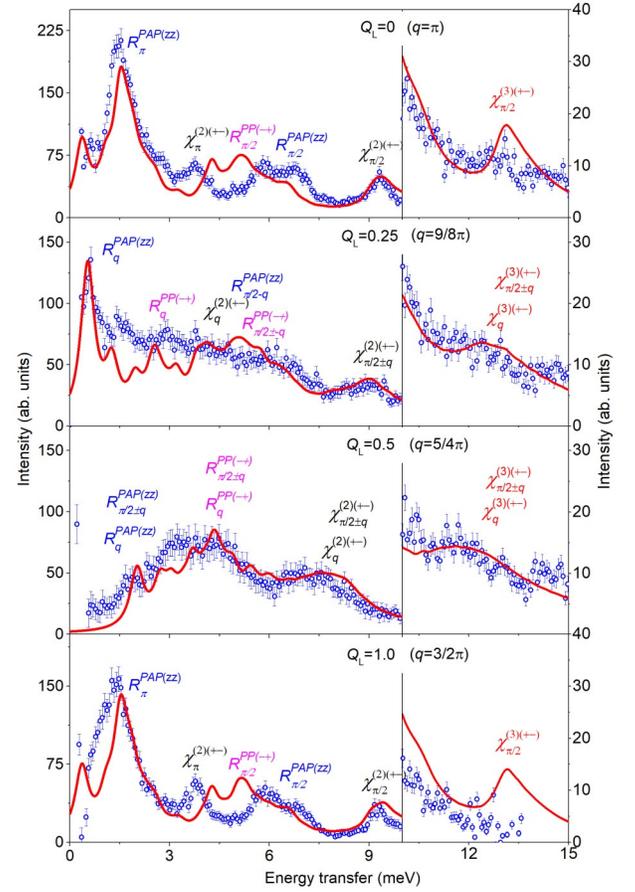

**Figure 3| Comparison of the scattering intensity at $B = 9$ T with Bethe Ansalz calculations.** Energy cuts through the data measured in the LET spectrometer at constant wave vectors $Q_L = 0$, 0.25, 0.5 and 1 r.l.u. ($q = \pi$, $9/8\pi$, $5/4\pi$, and $3/2\pi$; circular points), obtained from Fig. **2d** (data points over 0-10 meV) and **2f** (data points over 10-15 meV). The cuts are summed over the range $\Delta Q_L = \pm 0.1$ r.l.u. centered at the $Q_L$ values shown in the figures. The error bars indicate standard deviations assuming Poisson counting statistics. . For $Q_L=1.0$ at high energies (> 9 meV) the data are collected from detectors close to the edge of the detector bank which results into lower intensity. The red solid lines are the theoretical intensities obtained from the Bethe ansatz calculations (Fig. **2e**) by taking similar cuts. All the modes are labeled with the $q$ positions and the corresponding DSF channels (zz for $D^{zz}$; +− and −+ for $D^{+-}$ and $D^{-+}$, respectively).

Figure 4 depicts the field-dependent INS spectra measured at two reciprocal lattice points (3.25,0,0) and (2.3,2.3,0). In addition to the zone folding, the structure factor alters the intensities of the various modes from one Brillouin zone to another[14], providing an opportunity to investigate individual modes in particular detail. For example, only the high energy modes ($E> 4$ meV at 14.9 T) have intensity at (3.25,0,0), whereas, the low energy modes are more visible at (2.3,2.3,0). Of particular interest is the 3-string mode ($\chi_{\pi/2}^{(3)}$) which in spite of its weak intensity is clearly visible at (3.25,0,0) over the field range $4 \leq B \leq 12$ T. The intensities of the modes are shown by energy scans at $B = 6$, 9, 12 and 14.9 T for the $Q$ position (3.25, 0, 0) in Fig. 4c. Despite the complexity of the spectrum, good agreement between experimental INS results and theory is achieved for all the field values. Figure 4d shows the spectrum at unprecedentedly high magnetic fields up to 25.9 T. Since each peak is measured under different conditions the intensities are not comparable. Nevertheless the peak positions indicate how the various modes including the 2-string mode, continue to develop with increasing field towards saturation. In agreement with the THz results, all the modes have linear field dependences with distinct slopes and characteristic energies, except the $R_{\pi/2}^{PAP(zz)}$ mode which is field independent. The experimental slopes ($\partial E/\partial B$) are represented in Figs. 4a and 4b by solid lines and have the fitted values ~ 0.26(1), 0.27(1), -0.30(2), 0.44(5) and 0.60(2) meV/Tesla for the $\chi_{\pi/2}^{(2)}$, $R_\pi^{PAP}$, $R_{\pi/2}^{PP}$, $\chi_\pi^{(2)}$, and $\chi_{\pi/2}^{(3)}$, respectively. These results are in excellent agreement with the field dependences obtained from the Bethe Ansatz calculation (dashed lines in Figs. 4a and 4b; and Table S1[14]). Small differences in the absolute values are discernible, which may be due to the interchain exchange interactions in $SrCo_2V_2O_8$.

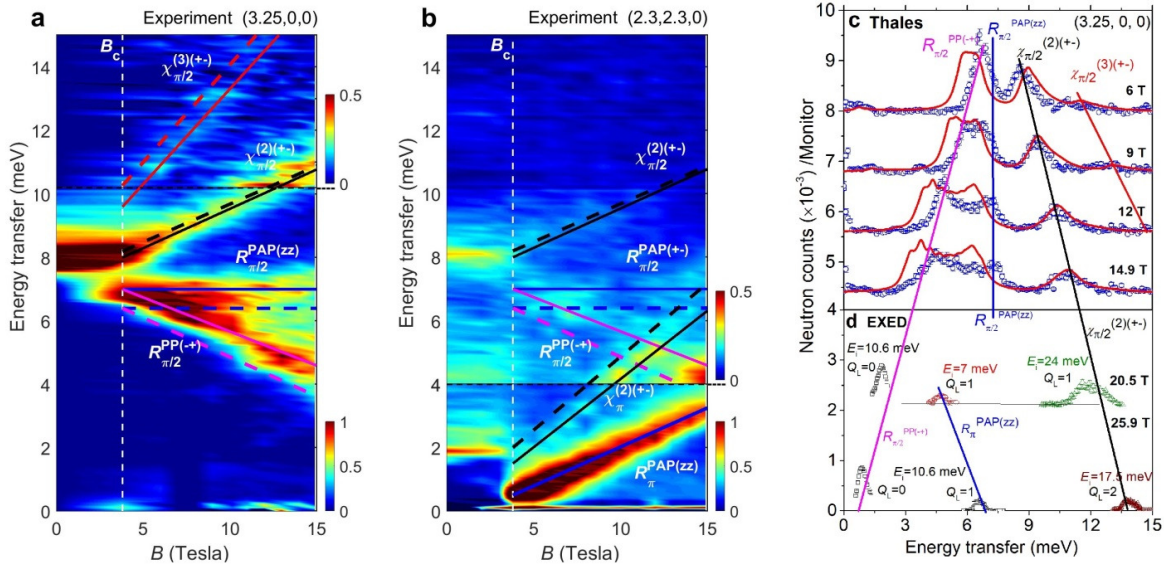

**Figure 4| Magnetic field dependence of Bethe strings.** False color maps of the inelastic neutron scattering (INS) intensity measured on ThALES spectrometer, ILL, France, at T=1.7 K showing the field dependence of the excitation spectrum of $SrCo_2V_2O_8$ for longitudinal magnetic fields up to 14.9 T for two selected $Q$-positions **a**, (3.25, 0, 0) and **b**, (2.3, 2.3, 0), respectively. The labels indicate the observed modes along with the wavevector $q$ and DSF channel where they appear. The solid coloured lines are guides to eye highlighting the field-dependences of the modes, while, the dashed lines are field dependence of the theoretical eigenenergies obtained from the Bethe ansaltz calculations. Despite small overall shifts in the energies of the measured modes compared to the Bethe ansatz calculations (for $R_\pi^{PAP(zz)}$ both the experimental and theoretical curves are overlapped), the field dependences of the modes agree very well with the theory. The experiment [theoretical] values of the field-dependences are 0.26(1) [0.27(1)], 0.27(1) [0.27(1)], -0.30(2) [-0.27(1)], 0.44(5) [0.50(2)] and 0.60(2) [0.63(3)] meV/Tesla for the $\chi_{\pi/2}^{(2)}$, $R_\pi^{PAP}$, $R_{\pi/2}^{PP}$, $\chi_\pi^{(2)}$, and $\chi_{\pi/2}^{(3)}$, respectively. **c**, The constant-$Q$ energy scans at $B=$ 6, 9, 12 and 14.9 T for (3.25, 0, 0). The error bars represent standard deviations. The intensities calculated by the Bethe ansatz method are shown by the solid red lines. The contributions of the screw chain structure factor of $SrCo_2V_2O_8$, the $Co^{2+}$ form factor, the anisotropic Lande g-factor, and the polarization factors have been considered for the comparison with the experimental data. **d**, The excitation spectra for longitudinal fields 20.5 and 25.9 T measured on HFM-EXED spectrometer, HZB, Germany, at $T$=1.3 K. Due to the instrumental configuration each peak was measured at different values of incident energy $E_i$ and wavevector $Q_L$ as shown by the labels, in each case an integration of $\Delta Q_L = \pm 0.2$ was made around the nominal $Q_L$-value.

The experimental data and theoretical calculations presented here provide a detailed study of the long-sought-after Bethe string excitations over the full Brillouin zone in unprecedented detail. The many-body Bethe strings are unambiguously identified in our study, based on the accurate agreement between theory and experiment in terms of their eigenenergies and scattering intensity and linewidths. This study may open up the opportunity to investigate magnetic analogues of the string-theory scenarios[8,29].

**Acknowledgments**

We acknowledge the HFM/EXED team and P. Smeibidl, R. Wahle and S. Gerischer for their technical support during the measurements. JW acknowledges support from Shanghai City. The high field experiments at Dresden were supported by HLD at HZDR, member of the European Magnetic Field Laboratory (EMFL).

**Author Contributions**

A.K.B. and B.L. conceived the experiments and coordinated the project. A.K.B. and A.T.M.N.I. prepared and characterized the high-quality single crystals. A.K.B., B.K. and J.M.L. performed the bulk measurements. A.K.B., B.L. and R. B. performed the LET experiments. B.L. and M. B. performed ThALES experiments. B. L., M. B., and O. P. performed the HFM/EXED measurements. A. K. B and B. L. analyzed all the data. J.W. and W.Y. carried out the Bethe-ansatz calculations. A.K.B. wrote the manuscript with contributions from B.L., J.W., W.Y. and Z.W. All authors discussed the data and its interpretation.

Supplementary Information:

# Dispersions of Many-Body Bethe strings

Anup Kumar Bera[1,2,*], Jianda Wu[3,*], Wang Yang[4,*], Zhe Wang[5], Robert Bewley,[6] Martin Boehm,[7] Maciej Bartkowiak,[1] Oleksandr Prokhnenko,[1] Bastian Klemke,[1] A. T. M. Nazmul Islam,[1] Joseph Mathew Law,[8] Bella Lake[1,9,*]

[1] Helmholtz-Zentrum Berlin für Materialien und Energie, 14109 Berlin, Germany
[2] Solid State Physics Division, Bhabha Atomic Research Centre, Mumbai 400085, India
[3] Tsung-Dao Lee Institute&School of Physic and Astronomy, Shanghai Jiao Tong University, 800 Dongchuan Rd., Shanghai, 200240, China
[4] Stewart Blusson Quantum Matter Institute, 2355 East Mall Vancouver, BC, V6T 1Z4, Canada
[5] Institute of Physics II, University of Cologne, 50937 Cologne, Germany
[6] ISIS Facility, STFC Rutherford Appleton Laboratory, Harwell Oxford, Didcot OX11 0QX, United Kingdom
[7] Institut Laue-Langevin, 71 Avenue des Martyrs, 38042 Grenoble, France
[8] Hochfeld-Magnetlabor Dresden (HLD-EMFL), Helmholtz-Zentrum Dresden-Rossendorf, D-01314 Dresden, Germany
[9] Institut für Festkörperphysik, Technische Universität Berlin, 10623 Berlin, Germany

## Methods

### Sample preparation and characterization

Good quality large single crystals of $SrCo_2V_2O_8$ (~ 4 mm diameter and 100 mm long) were grown (Fig. S1a) by the floating zone method[30] using an optical floating zone furnace (Crystal Systems, Japan)[30] at the Core Laboratory for Quantum Materials (QM core lab), Helmholtz Zentrum Berlin, Germany. Initially, high purity powder samples of $SrCo_2V_2O_8$ were prepared by the solid state reaction method from high-purity (> 99.99%) reagents of $SrCO_3$, $CoC_2O_4 \cdot 2H_2O$, and $V_2O_5$ at 800 °C in air. A cylindrical feed rod was prepared from the powder samples, which was pressed hydrostatically up to 3,000 bar in a cold isostatic press and sintered in air at 800 °C for 12 h. The floating zone furnace was used to melt the tip of the feed rod, which recrystallized on a seed crystal to achieve single-phase growth. A stable growth was achieved in a mixture of high-quality flowing oxygen (20%) and argon (80%) at ambient pressure with a growth rate of 1 mm/h. The crystal quality was checked by x-ray Laue diffraction (Fig. S1b), as well as by powder x-ray diffraction of the ground single crystal using a laboratory x-ray machine (Bruker D8 Advance).

The temperature dependent magnetic susceptibility (Fig. S2a) was measured on a cut single crystal (dimension ~ 2 × 2 × 1 mm$^3$) over the temperature range of 2 - 900 K using a Quantum Design physical property measurement system (PPMS) at the QM core Lab. The susceptibility curves show a magnetic transition at $T_N$ = 5 K (see Fig S2a) with a broad peak centered around 30 K revealing 1D short-range spin-spin correlations[15]. A difference between parallel and perpendicular susceptibilities up to a temperature of ~ 800 K indicates the presence of a large paramagnetic anisotropy. The inverse susceptibility ($1/\chi$) curves measured with longitudinal and transverse fields (Fig. S2b) show a non-linear behavior below 450 K indicating the presence of low-lying crystal field excitations of $Co^{2+}$ ion in a $CoO_6$ octahedral environment that appears at ~ 30 meV. The magnetic moment of the $Co^{2+}$ ions in the distorted octahedral crystal field environment is described well by a highly anisotropic *pseudospin*, $S = 1/2$[31]. In the magnetically ordered state below 5 K, the lower values of magnetization for the field applied parallel to the chain axis (c axis) yields that the crystallographic c axis is the magnetic easy axis which is further confirmed from the high field magnetization study.

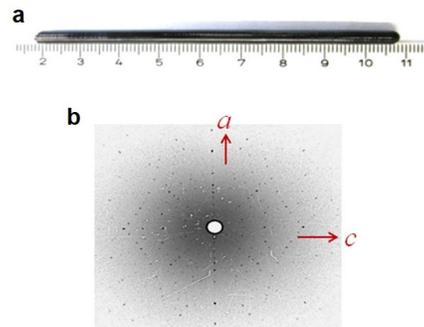

**Figure S1| a, Single crystal and Laue pattern**. The single crystal of $SrCo_2V_2O_8$ grown by the floating zone method. **b**, The x-ray Laue backscattering diffraction pattern from the crystal used for neutron scattering study.

Further, heat capacity and magnetocaloric effect measurements were carried out on a cut single crystal (2 × 2 × 1 mm$^3$) in the QM Core Lab using a cryogenic system consisting of an Oxford

Instruments 14.5 T cryomagnet and a Heliox $^3$He insert. With this in-house developed experimental setup, the caloric measurements have been performed up to magnetic field of 14.5 T and down to temperatures of 0.3 K. The heat capacity data were obtained using the standard pulse relaxation method, the quasi-adiabatic pulse method, and the dual-slope method.

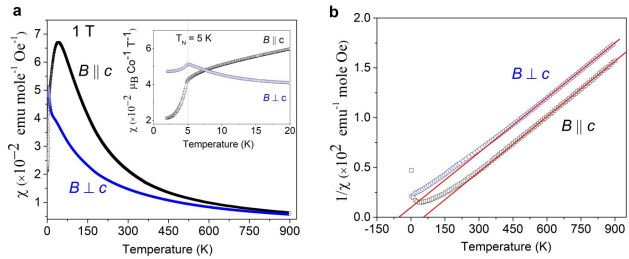

**Figure S2| Magnetic susceptibility of $SrCo_2V_2O_8$. a,** The temperature-dependent susceptibility ($\chi$) curves for $SrCo_2V_2O_8$, measured under 1 Tesla of magnetic field, for the field applied parallel to the chain direction ($B \parallel c$) and perpendicular to the chain direction ($B \perp c$), respectively. The insets show the enlarged view of the $\chi(T)$ curves over the temperature range 0-20 K. The magnetic transition is illustrated by the clear anomalies in both susceptibility curves. **b,** The inverse susceptibility curves as a function of temperature. The solid lines are the linear fits to the experimental data. The data over the temperature range 450-900 K was used for the fittings.

The high-field magnetization measurements were performed by using a 60 T pulsed-field magnet (pulse duration of 25 ms) at the Hochfeld Magnetlabor Dresden (HLD), Helmholtz Zentrum Dresden Rossendorf, Germany. The magnetization signal was detected by an induction method with a standard pick-up coil system[32]. The sample magnetization was determined after subtraction of an empty magnetometer background.

Magnetic excitations in $SrCo_2V_2O_8$ under longitudinal magnetic field were mapped through inelastic neutron scattering (INS) measurements performed on a number of instruments. For all the INS measurements the same single crystal (weight ~ 4.5 g, diameter ~4 mm, and length ~40 mm)) was used in a longitudinal magnetic field.

Time-of-flight neutron scattering measurements were carried out on the cold neutron multi-chopper LET spectrometer at the ISIS facility of the Rutherford Appleton Laboratory (RAL), UK. Measurements on LET were performed with a fixed incident neutron energy of $E_i$ = 20 meV with the repetition rate multiplication method[33,34] by using a straight Gd Fermi chopper which provides simultaneous measurement of INS patterns corresponding to the incident energy of $E_i$ =20, 6, 2.8 and 1.6 meV, respectively. The choppers speeds were fixed to 200 and 100 Hz for the $T_0$ chopper and Fermi chopper, respectively. The sample was fixed on a Cu holder and mounted in a 9 T vertical field cryomagnet with a dilution refrigerator insert. The sample was mounted with the c axis vertical and (h,0,0) initially aligned along the incident neutron wave vector $k_i$. Scattering patterns were recorded at $T$= 0.08 K under $B$=0, 3, 6, and 9 T of vertical magnetic fields parallel to the c axis. Additional pattern was recorded at 2.5 K under 9 T of vertical magnetic field. To map the complete dynamic structure factor $S$ ($q$, $\omega$) the sample was rotated through 60° in 1° steps and measured for a fixed amount of proton charge on the spallation target (roughly 10 minutes per measurement). The individual data of each rotation angle were normalized for the proton charge on the spallation target, corrected for detector sensitivity using vanadium normalization, and then binned from 4D laboratory coordinates to 4D sample coordinates using standard direct geometry chopper spectrometer reduction routines within the Mantid software[35].

Another set of measurements were conducted on the ThALES cold triple-axis spectrometer at ILL, Grenoble, France. Here the sample was also fixed on a Cu holder with the c axis vertical and put inside a 15 T vertical Cryo magnet. Measurements were performed over the magnetic field range 0–14.9 T and at a base temperature of 1.7 K. Data were collected with a fixed final energy of $E_f$=3.5meV ($k_f$= 1.3 Å$^{-1}$) using both the double-focusing PG monochromator and PG analyzer. A velocity selector was used to filter out the higher-order neutrons. The energy resolution was 0.11 meV at the elastic line. Measurements with an applied field on ThALES were confined to the (h,k,0) scattering plane.

Further measurements of magnetic excitations were carried out at the HFM/EXED high magnetic field facility for neutron scattering which can achieve DC magnetic fields up to $B$= 25.9 T$^{36,37}$. The sample was cooled to $T$=1.3 K using a $^3$He cryostat and was oriented with the crystallographic c-axis parallel to the horizontal magnetic field. The measurements were carried out at 3 magnetic field values, 15, 20.5 and 25.9 T. The instrument was used in direct time-of-flight spectroscopy mode with a monochromatic chopper placed in front of the sample[37]. The detector geometry ($\Delta 2\theta$=30°) combined with the sample orientation restricted the reciprocal space coverage along the chains (*c*-direction). Therefore in order to target the various excitations different incident energies were used to achieve the required energy transfers at different reciprocal lattice positions for *L*=0, 1 and 2. Because different incident energies were used the peaks were measured with different resolution values and overall intensities thus the intensities of the peaks cannot be easily modeled, nevertheless this measurement gives accurate values for the peak positions and extends our phase diagram to much higher longitudinal fields close to the saturation field $B_S$.

Spin dynamics of the spin 1/2 XXZ antiferromagnetic chain SrCo$_2$V$_2$O$_8$ were calculated by algebraic Bethe ansatz formalism[26-28], and dynamic spin structure factors of various excitations were estimated under the guidance of sum rules. In the algebraic Bethe ansatz formalism, the matrix elements of local spin operators between two different Bethe eigenstates are expressed in terms of the determinant formulae in finite systems[38,39].

**Crystal structure, real and reciprocal space definitions**

The crystal structure of the studied spin-1/2 XXZ chain compound SrCo$_2$V$_2$O$_8$ is described by tetragonal space group *I4$_1$cd* with lattice constants $a$ = 12.2420(1) Å and $c$ = 8.4086(1) Å at 2 K. The only magnetic ion Co$^{2+}$ in SrCo$_2$V$_2$O$_8$ is located at the Wyckoff site 16*b*. The screw chains in SrCo$_2$V$_2$O$_8$ are formed by edge-shared CoO$_6$ octahedra and running along the *c* axis with a period of four Co$^{2+}$ ions per unit cell (Fig. S3a). There are four such screw chains per unit cell centered around (1/4,1/4), (1/4,3/4), (3/4,1/4), and (3/4,3/4) in the *ab* plane (Fig. S3). Among them, one pair of diagonal chains rotate clockwise, while the other pair rotate counterclockwise when propagating along the *c* axis. The fourfold periodicity of the screw chains along with the fact that each of the magnetic ions is shifted from the center of the chain axis results in six excitation modes with different structure factors in the reciprocal plane as per the following equations

$$S(q,\omega) = \cos^2\psi_1 \cos^2\psi_2\, S'(h,k,l)$$
$$+ \frac{\cos^2\psi_1 \sin^2\psi_2}{2}[S'(h+1,k,l+1)$$
$$+S'(h+1,k,l+3)]$$
$$+ \frac{\sin^2\psi_1 \cos^2\psi_2}{2}[S'(h,k+1,l+1)$$
$$+S'(h,k+1,l+3)]$$
$$+ \sin^2\psi_1 \sin^2\psi_2\, S'(h,k,l)$$

Where $S'(h,k,l)$ is the structure factor of a single straight chain, $\psi_1 = \frac{2\pi d}{a}h$ and $\psi_2 = \frac{2\pi d}{a}k$, and *d*

(~0.08*a*) is the offset of each Co$^{2+}$ ion with respect to the central axis [at (1/4,1/4) or (1/4,3/4) or (3/4,1/4) or (3/4,3/4)] of the corresponding screw chains. Each of the individual modes has a periodicity of 4 reciprocal lattice units (r.l.u.) along the $Q_L$ in reciprocal space as there are four Co$^{2+}$ ions per chain per unit cell in real space. Due to the fact that there are four equivalent screw chains per unit cell each with four Co$^{2+}$ ions per *c*-lattice parameter a total of six "copies" of the cross-section for a single chain, which are shifted with respect to one another by one r.l.u. along the chain direction [for details see [40]]. The excitation modes are shifted in the reciprocal space such that each of the reciprocal lattice points (any integer combinations of *h*, *k*, and *l*) becomes an AFM zone center. As the structure factor of individual modes is decided by the momentum transfer perpendicular to the chain axis (values of *h* and *k*), a substantial variation of relative intensity of the modes appears for the measured inelastic neutron scattering spectra of SrCo$_2$V$_2$O$_8$. The screw chains of CoO$_6$ are connected by edge-sharing VO$_4$ (V$^{5+}$; 3$d^0$, *S* = 0) tetrahedral which renders weak ($J^\perp_{eff}/J < 10^{-2}$) but finite interchain superexchange interactions expected to have both in-plane as well as out-of-plane components as found by us for the isostructural compound SrNi$_2$V$_2$O$_8$[40]. The weak $J^\perp$ leads to a magnetic long-range ordering below $T_N$ = 5 K in zero field, and is also responsible for the stabilization of long-range magnetic ordering in the critical regime (*B*> $B_{c1}$) at lower temperatures (*T* < 0.8 K).

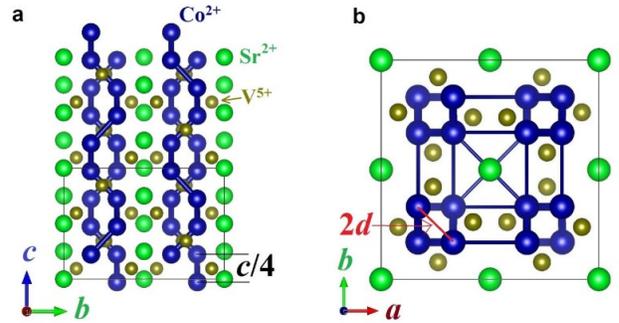

**Figure S3|** **a,** The four-fold screw chains along the *c* axis. The O$^{2-}$ ions are absent for clarity. The solid thick lines are representation of the nearest-neighbour intrachain interaction *J*. The unit cell dimensions are shown by the gray lines. **b,** Projection of the screw chains onto the *ab* plane. There are four screw chains per unit cell centered at (1/4,1/4), (1/4,3/4), (3/4,1/4), (3/4,3/4). Each of the magnetic Co$^{2+}$ ions is shifted by an amount of *d* from their corresponding central axis. The couplings between chains are shown by the thin lines.

**Zero field spin excitation spectrum**

The spin-1/2 XXZ chain model described in Eq. 1 possesses spinon excitations (in zero field) which are particles with fractional spin ($S=1/2$) and are fundamentally different from the magnons or spin-wave excitations (spin $S=1$) of conventional magnets observed in higher dimensional structures. A magnon can be viewed as a pair of spinons bound together or '*confined*'. In one-dimension, however, they can unbind and become free spinons observed as a spinon continuum – a process known as deconfinement (coupling between the spin chains stabilizes long-range magnetic order at a finite temperature ($T > 0$), and allows the process of spinon confinement to be directly observed, with analogies to quark confinement in quantum chromodynamics. The zero-field spin excitation spectra of $SrCo_2V_2O_8$ in the magnetic ordered state (at 1.7 K) below the Néel temperature $T_N$ = 5 K, measured by inelastic neutron scattering, are shown in Fig. S4. The excitation spectra, reveal discrete excitations consisting of 2-spinons bound states[16] confined by weak interchain couplings. And, therefore, the magnetic state of $SrCo_2V_2O_8$ is not a true classical Néel state as it lacks conventional spin wave excitations. The observed bound modes are strongly dispersive along the *c* axis (and weakly dispersive along the *a* and *b* directions[16]) (Fig. S4). As depicted in Fig. S4c, an energy scan at (2.3,2.3,0) reveals a sequence of discrete, resolution-limited peaks at 1.7 K. Each of the peaks consists of two closely spaced peaks due to the parity of the number *N* of flipped spins between two domain walls. The modes with *N* equal to odd and even carry a spin $S^z = \pm 1$ and 0, respectively[16,22]. In a neutron scattering experiment, the modes correspond to $S^z = \pm 1$ appear in the transverse (T-mode) excitations (spin fluctuations in the plane perpendicular to the direction of the ordered magnetic moment which is along the *c* axis). On the other hand, the modes correspond to $S^z = 0$ emerge as longitudinal (L-mode) excitations (spin fluctuations parallel to the ordered moment). The overall excitation spectrum is well accounted by a $S=1/2$ XXZ chain model with $J \approx 7.0$ meV and $\Delta \approx 2.0$, respectively. An additional broad peak around 8 meV appears due to the zone folding effect which leads to overlapping excitation modes whose zone centres are separated by one r.l.u.

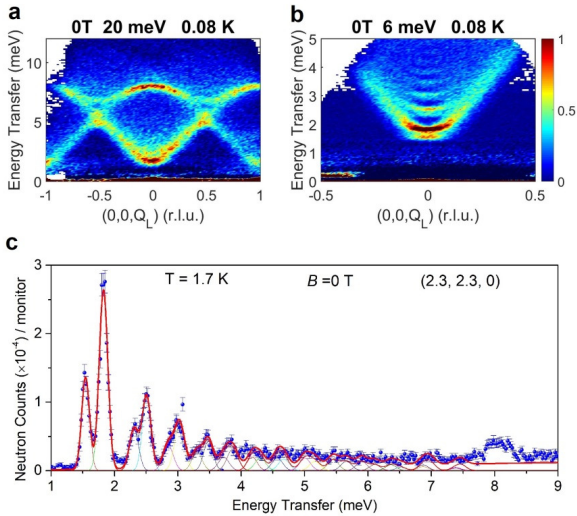

**Figure S4| Magnetic excitation spectrum of $SrCo_2V_2O_8$ in zero field**. **a-b**, The measured inelastic neutron scattering spectra on LET spectrometer, ISIS, UK, at 0.08 K with an incident energy $E_i$ = 20 and $E_i$ =6 meV., respectively. To gain intensity, the data was integrated over $Q_H = Q_K$ = -3.5–0 r.l.u for $E_i$ =20 meV and $Q_H = Q_K$= -3.5–-1.5 for $E_i$ =6 meV, respectively. The intensities are denoted by colors. **c**, The measured energy scan at $Q$= (2.3, 2.3, 0) by Thales spectrometer, ILL, France, at 1.7 K. Error bars indicate standard deviations assuming Poisson counting statistics. A sequence of pairs of peaks is evident with energy. The pair of peaks appear due to the longitudinal and transverse polarizations. The red curve is a fit of Gaussian peaks to the data. Individual peaks are shown by the thin lines.

**Magnetic phase diagram of $SrCo_2V_2O_8$**

In zero field, our neutron diffraction study reveals collinear Néel antiferromagnet (NAF) ordering below the $T_N$ =5 K with the spins pointing parallel to the *c* axis (chain axis) and a parallel/antiparallel arrangement of the spins along the *a/b* directions[15]. In the Néel state, the antiparallel alignment of the spins within the chains corresponds to a total spin-z quantum number $S_Z^T$=0. In a longitudinal applied magnetic field (along the magnetically easy *c* axis) $SrCo_2V_2O_8$ shows a field-induced transition at $B_{c1}$ = 3.8 T from the 3D long-range Néel state to a 1D quantum critical Tomonaga-Luttinger liquid state[41] before reaching a fully field-polarized state at $B_s$ = 28.7 T [Fig. 1B]. At lower temperatures ($T < 1$ K), long-range magnetic orderings with distinct characteristics are stabilized in the quantum critical regime by the weak three-dimensional couplings between the spin chains (Fig. S5). The ordered states are identified as incommensurate longitudinal spin density wave (LSDW) ($B_{c1} < B < B_{c2}$=7.2 T) and a transverse antiferromagnet (TAF) $B > B_{c2}$, respectively[42,43]. All the phase transitions between the order and disorder states are found to be second-order except the low-temperature order–order transition between NAF and LSDW which is of the first order. Such a phase diagram was also reported by other groups for the studied compound $SrCo_2V_2O_8$[41] as well as for the related compound[43,44].

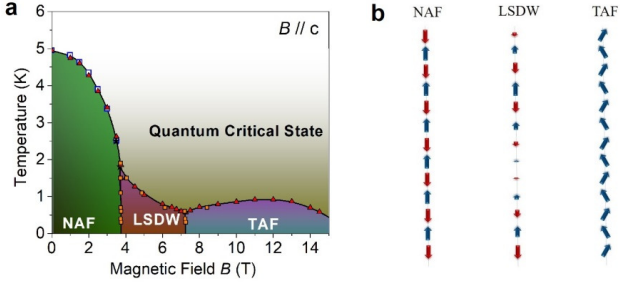

**Figure S5| Magnetic phase diagram of $SrCo_2V_2O_8$ in the *B-T* plane. a**, The magnetic phase diagram in the magnetic field and temperature plane. The phase boundaries are obtained from the temperature dependent specific heat (triangles), field-dependent magnetocaloric effect (filled squares), and temperature (open squares) and magnetic field (stars) dependent dc magnetization studies. Weak interchain couplings stabilize ordered magnetic states at low temperatures. NAF, LSDW, and TAF stand for Néel antiferromagnetic state, longitudinal spin density wave state, and transverse antiferromagnetic state, respectively. **b**, The schematic representation of the NAF, LSDW, and TAF states.

## Bethe ansatz formalism

The Bethe ansatz method is employed to obtain the eigenstates of the periodic antiferromagnetic spin-1/2 Heisenberg-Ising chain described by the following Hamiltonian,

$$H(\Delta, h) = J \sum_{n=1}^{N} \left[ \left( S_n^x S_{n+1}^x + S_n^y S_{n+1}^y \right) + \Delta S_n^z S_{n+1}^z \right] - h_z \sum_{n=1}^{N} S_n^z,$$

where the nearest-neighbor coupling $J > 0$, the Ising anisotropic parameter $\Delta = \cosh \eta > 1$, and the longitudinal field $h_z = g_z \mu_B B$ with $g_z$, $\mu_B$, and $B$ being the Lande factor, Bohr magneton, and external magnetic field, respectively. The system is Neel ordered at zero field, and becomes gapless above critical field $h_c = J \sinh \eta \sum_{l=-\infty}^{+\infty} \frac{(-)^l}{\cosh n\eta}$ until saturation field $h_s = J(\Delta + 1)$ [28,45].

Regarding up-spins as vacuum and down-spins as particles, a Bethe eigenstate with total spin $S_T^z = \sum_{n1}^{N} S_n^z = N/2 - r$ is characterized by $r$ rapidities $\{\lambda_j\}_{1 \leq j \leq r}$, which are related to the particles' momenta $\{k_j\}$ and energies $\{\epsilon_j\}$ through $e^{ik_j} = \frac{\sin\left(\lambda_j + \frac{i\eta}{2}\right)}{\sin\left(\lambda_j - \frac{i\eta}{2}\right)}$ and $\epsilon_j = J(\cos k_j - \Delta)$. An eigenstate is called a real Bethe eigenstate if all rapidities are real or a string state if there exist complex-valued rapidities. The rapidities can be determined by solving the following Bethe ansatz equations (BAEs)[46]

$$N\theta_1(\lambda_j) = 2\pi I_j + \sum_{l=1}^{r} \theta_2(\lambda_j - \lambda_l), \quad j = 1, \dots, r,$$

in which $\{I_j\}_{1 \leq j \leq r}$ are Bethe quantum numbers (BQNs) which are integers when $r$ is odd and half-odd integers when $r$ is even, and the functions $\theta_1$ and $\theta_2$ are defined by

$$\theta_j(\lambda) \equiv 2 \arctan\left(\frac{\tan \lambda}{\tanh\left(\frac{j\eta}{2}\right)}\right) + 2\pi \left\lfloor \frac{\text{Re}(\lambda)}{\pi} + \frac{1}{2} \right\rfloor, \quad j = 1,2,$$

with $\lfloor A \rfloor$ denoting the floor function which gives the largest integer not greater than $A$.

The BQNs of the ground state are given by either all integers (when $r$ is even) or all half-odd integers ($r$ odd) within the range $[-\frac{r-1}{2}, \frac{r-1}{2}]$ [28]. Either the psinon-psinon states, or the psinon-antipsinon states can be used to classify the real Bethe eigenstates[47]. The BQNs of an $m$-pair psinon-psinon state ($m\psi\psi$) can be obtained by choosing $r$ numbers within the range $[-\frac{r-1}{2} - m, \frac{r-1}{2} + m]$, while an $n$-pair psinon-antipsinon state ($m\psi\psi^*$) is obtained by choosing $r - m$ numbers within the range $[-\frac{r-1}{2}, \frac{r-1}{2}]$ and $m$ numbers outside this range.

Next, we turn to string states. The rapidities of an $n$-string are given by

$$\lambda_j^n = \lambda^{(n)} + i(n + 1 - 2j)\frac{\eta}{2} + \delta_j^n, \quad 1 \leq j \leq n,$$

in which the real number $\lambda^{(n)}$ is the string center, and $\{\delta_j^n\}$ parameterize the string deviations. Within the string ansatz, the string deviations are neglected when $N$ is large, and the BAEs can be simplified accordingly[46]. For a string state consisting of $r_n$ $n$-strings with $\sum_n n r_n = r$, the string centers are determined by solving the following reduced Bethe-Gaudin-Takahashi equations[46]

$$N\theta_n(\lambda_\alpha^{(n)}) = 2\pi I_\alpha^{(n)} + \sum_{(m\beta) \neq (n\alpha)} \Theta_{nm}(\lambda_\alpha^{(n)} - \lambda_\beta^{(m)}),$$
$$1 \leq \alpha \leq r_n, 1 \leq \beta \leq r_m,$$

in which $I_\alpha^{(n)}$ are the reduced BQNs, and the function $\Theta_{nm}$ is defined as

$$\Theta_{nm} = (1 - \delta_{nm})\theta_{|n-m|} + 2\theta_{|n-m|+2} + \cdots + 2\theta_{n+m-2} + \theta_{|n+m|}.$$

Here we list the rules for the reduced BQNs of string states which contain a single $n$-string ($n \geq 2$) with other rapidities being real[48]. For the string state $1\chi^{(n)}m\psi\psi$ ($1\chi^{(n)}m\psi\psi^*$), the reduced BQN $I^{(n)}$ of $\chi^{(n)}$ lies within the range $\left[-\frac{N-2r}{2}, \frac{N-2r}{2}+2n-1\right]$, while the BQNs of $m\psi\psi$ ($m\psi\psi^*$) are obtained from those of real Bethe eigenstates by replacing $r$ with $r-n$.

### Determinant formulas

The determinant formulas have been developed to efficiently calculate spin form factors in the basis of Bethe eigenstates for rather large system size[48]. For example, defining $S_q^a = \frac{1}{\sqrt{N}}\sum_j e^{iqj}S_j^a$ ($a = x, y, z$), the matrix element of $S_q^-$ between two Bethe eigenstates with rapidities $\{\mu_j\}_{1\leq j\leq M+1}$ and $\{\lambda_j\}_{1\leq j\leq M}$ is given by

$|\langle\{\mu\}|S_q^-|\{\lambda\}\rangle|^2 =$

$N\delta_{k(\{\lambda\})-k(\{\mu\}),q}|\sin i\eta|\frac{\prod_{k=1}^{M+1}\left|\sin\left(\mu_k-\frac{i\eta}{2}\right)\right|^2}{\prod_{j=1}^{M}\left|\sin\left(\lambda_j-\frac{i\eta}{2}\right)\right|^2} \times$

$\frac{1}{\prod_{k\neq k'}|\sin(\mu_k-\mu_{k'}+i\eta)|\prod_{j\neq j'}|\sin(\lambda_j-\lambda_{j'}+i\eta)|}\frac{|\det H^-|^2}{|\det\Phi(\{\mu\})||\det\Phi(\{\lambda\})|}$

where the $(M+1)\times(M+1)$ matrix $H^-$ follows

$H_{kj}^- = \frac{1}{\sin(\mu_k-\lambda_j)}[\prod_{l=1(l\neq k)}^{M+1}\sin(\mu_l-\lambda_j+i\eta)-\left(\frac{\sin(\lambda_j-i\eta/2)}{\sin(\lambda_j+i\eta/2)}\right)^N\prod_{l=1(l\neq k)}^{M+1}\sin(\mu_l-\lambda_j-i\eta)]$, $1\leq k\leq M+1, 1\leq j\leq M$,

and

$H_{k,M+1}^- = \frac{1}{\sin\left(\mu_k+\frac{i\eta}{2}\right)\sin(\mu_k-i\eta/2)}$, $1\leq k\leq M+1$.

Other matrix elements have similar expressions[48]. Within the string ansatz approach, if the matrix element involves string states, the determinant formulas need to be regularized[48].

### Theoretical magnetization versus magnetic field

Define $H_0$ as the Hamiltonian without magnetic field,

$$H_0(\Delta) = J\sum_{n=1}^{N}\left[\left(S_n^x S_{n+1}^x + S_n^y S_{n+1}^y\right) + \Delta S_n^z S_{n+1}^z\right],$$

and $m$ as the magnetization per site,

$$m = \frac{1}{N}\langle G|S_T^z|G\rangle,$$

where $|G\rangle$ is the ground state, then $h$ and $m$ are related by the Legendre transform as

$$h = \frac{\partial e_0}{\partial m},$$

where $e_0 = \frac{1}{N}\langle G|H_0|G\rangle$ is the energy density.

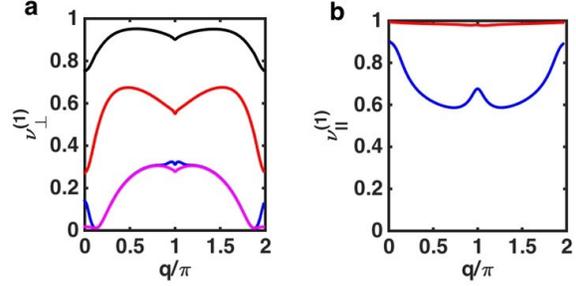

**Figure S6|** **a**, $\nu_\perp^{(1)}$ and **b**, $\nu_\parallel^{(1)}$ at $2m = 12\%$. The system size is taken as $N = 100$ within numerical computations. In (**a**), the pink, blue, red, and black curves represent the cumulative sum ratios by successively including $n\psi\psi$ ($n = 1, 2$) states in $D^{-+}$, $n\psi\psi^*$, $1\chi^{(2)}R$ and $1\chi^{(3)}R$ ($R = n\psi\psi^*$, $n\psi\psi, n = 1, 2$) states in $D^{+-}$, respectively. In (**b**), the blue and red curves represent the cumulative sum ratios by including $n\psi\psi^*$ and $1\chi^{(2)}R$ ($R = n\psi\psi^*$, $n\psi\psi, n = 1,2$) states in $D^{zz}$.

### Dynamical structure factors

The zero temperature dynamical structure factors (DSFs) are defined as

$$D^{a\bar{a}}(q,\omega) = 2\pi\sum_\mu\left|\langle\mu|S_q^{\bar{a}}|G\rangle\right|^2\delta(\omega - E_\mu + E_G),$$

in which $\bar{a} = -a$ ($a = \pm$), $\bar{a} = z$ ($a = z$), $|G\rangle$ is the ground state, $|\mu\rangle$ is the intermediate excited state, and $S_q^\pm = \frac{1}{\sqrt{N}}\sum_n e^{iqn}S_n^\pm$.

### Momentum-resolved first frequency moment sum rules

The transverse first frequency moment (FFM) sum rule follows[49]

$W_\perp = \int_0^\infty\frac{d\omega}{2\pi}\omega[D^{+-}(q,\omega) + D^{-+}(q,\omega)] = \alpha_\perp + \beta_\perp\cos q,$

where $\alpha_\perp = -e_0 - \frac{\Delta\partial e_0}{\partial\Delta} + mh$ and $\beta_\perp = (2-\Delta^2)\frac{\partial\Delta}{\partial e_0} + \Delta e_0$. The longitudinal FFM sum rule is[49]

$$W_\parallel(q) = \int_0^\infty \frac{d\omega}{2\pi} \omega D^{zz}(q,\omega) = (1-\cos q)\alpha_\parallel,$$

in which $\alpha_\parallel = -e_0 + \Delta \frac{\partial e_0}{\partial \Delta}$. The level of saturation of these sum rules can be evaluated by taking the FFM sum ratios $\nu_\perp^{(1)} = \frac{\overline{W}_\perp(q)}{W_\perp(q)}$ and $\nu_\parallel^{(1)} = \frac{\overline{W}_\parallel(q)}{W_\parallel(q)}$, where $\overline{W}_{\perp(\parallel)}(q)$ are calculated by only including the selected excitations. The Fig S6 display the FFM sum ratios $\nu_\perp^{(1)}(q)$ and $\nu_\parallel^{(1)}(q)$ at a representative magnetization $2m = 12\%$ with Ising anisotropy $\Delta = 2$. Both the sum rules of transverse and longitudinal DSFs are saturated to an excellent level.

## Comparison between theoretical and experimental slopes

Bellow we present a comparison between the slopes ($dE/dB$) of the all experimentally observed modes $\chi_{\pi/2}^{(3)}$, $\chi_\pi^{(2)}$, $\chi_{\pi/2}^{(2)}$, $R_\pi^{PAP}$, $R_{\pi/2}^{PAP}$ and $R_{\pi/2}^{PP}$ over the field range 6 - 25.9 T. The experimental resonance energies for the modes at 6, 9, 12, 14.9, 20.5 and 25.9 T are compared to the theoretical eigenenergies obtained from the BA calculations for the reduced magnetization values $2m = 0.08, 0.12, 0.2, 0.24, 0.38,$ and $0.60$ (Table S1). This is to mention here that all the modes have linear field dependences with distinct slopes, except the $R_{\pi/2}^{PAP}$ mode which is field independent. An excellent agreement between the experimental and theoretical slopes are found, although, there is small difference in the absolute values of the experimental resonance energies and calculated eigenenergies for some of the modes. Such small differences may be due to the interchain exchange interactions which will slightly shift the energy positions of some excitations. Interestingly, the slopes of the 2-string mode at the momentum positions $q = \pi$ and $\pi/2$ are quite different (~0.47 and ~ 0.27 meV/Tesla, respectively). The experimentally accessible 2-string mode at the momentum positions $q = \pi$ and $\pi/2$ by neutron scattering allows us for such a comparison.

**Table S1:** Comparison between theoretical and experimental slopes of the $\chi_{\pi/2}^{(3)}$, $\chi_\pi^{(2)}$, $\chi_{\pi/2}^{(2)}$, $R_\pi^{PAP}$, and $R_{\pi/2}^{PP}$ modes.

| B (T) | Eigen energies (meV) | | | | | | | | | |
|---|---|---|---|---|---|---|---|---|---|---|
| | $\chi_{\pi/2}^{(3)}$ | | $\chi_\pi^{(2)}$ | | $\chi_{\pi/2}^{(2)}$ | | $R_\pi^{PAP}$ | | $R_{\pi/2}^{PP}$ | |
| | Exp. | Theory | Exp. | Theory | Exp. | Theory | Exp. | Theory | Exp. | Theory |
| 6 | 10.9 | 11.5 | 2.5 | 3.2 | 8.6 | 8.9 | 1.0 | 1.0 | 6.7 | 5.8 |
| 9 | 12.4 | 13.2 | 3.9 | 4.4 | 9.3 | 9.5 | 1.6 | 1.6 | 5.8 | 5.1 |
| 13 | 15.1 | 15.9 | 5.2 | 6.65 | 10.0 | 10.2 | 2.6 | 2.7 | 4.8 | 4.0 |
| 14.9 | | | 6.7 | 7.6 | 11 | 11.1 | 3.1 | 3.2 | 4.4 | 3.4 |
| 20.5 | | | | | 11.8 | 12.4 | 4.8 | 4.9 | 1.8 | 1.7 |
| 25.9 | | | | | 13.8 | 14.2 | 6.53 | 6.3 | 0.9 | 0.7 |
| Slope $dE/dB$ (meV/Tesla) | 0.60(2) | 0.63(3) | 0.44(5) | 0.50(2) | 0.26(1) | 0.27(1) | 0.27(1) | 0.27(1) | -0.30(2) | -0.27(1) |